\documentclass{article}

\usepackage[preprint,nonatbib]{neurips_2026}

\usepackage[utf8]{inputenc} 
\usepackage[T1]{fontenc}    
\usepackage{hyperref}       
\usepackage{url}            
\usepackage{booktabs}       
\usepackage{amsfonts}       
\usepackage{nicefrac}       
\usepackage{microtype}      
\usepackage{xcolor}         

\usepackage{amssymb}
\usepackage{xspace}
\usepackage{multirow}
\usepackage{grffile}
\usepackage{tabularx}
\usepackage{enumitem}

\renewcommand{\vec}[1]{\mathbf{#1}}

\newcommand{\knn}{$k$-NN\xspace}
\newcommand{\Deltar}{$\Delta r$\xspace}
\newcommand{\nsplit}{\ensuremath{n_\mathrm{split}}\xspace}
\newcommand{\vlin}{\ensuremath{\vec{v}^\mathrm{lin}}\xspace}
\newcommand{\vtrue}{\ensuremath{\vec{v}^\mathrm{true}}\xspace}
\newcommand{\vpred}{\ensuremath{\vec{v}^\mathrm{pred}}\xspace}
\newcommand{\ngal}{\ensuremath{{n_\mathrm{gal}}}\xspace}
\newcommand{\lmax}{\ensuremath{{L_\mathrm{max}}}\xspace}

\newcommand{\nhat}{\ensuremath{\hat\vec{n}}\xspace}

\newcommand{\CosmoBench}{\textsc{CosmoBench}\xspace}
\newcommand{\Quijote}{\textsc{Quijote}\xspace}
\newcommand{\flamingo}{\textsc{FLAMINGO}\xspace}

\newcommand{\Velformer}{\textsc{Velocityformer}\xspace}

\newcommand{\hMpc}{\,h^{-1}\mathrm{Mpc}}

\newcommand{\ihMpcV}{\,h^3\mathrm{Mpc}^{-3}}
\newcommand{\hGpcV}{\,h^{-3}\mathrm{Gpc}^3}

\usepackage{environ}
\NewEnviron{splitequation}{%
    \begin{equation}\begin{split}
        \BODY
    \end{split}\end{equation}}

\usepackage{aas_macros}

\usepackage[style=authoryear,
            natbib=true,
            hyperref=true,
            url=false,        
            doi=false,         
            isbn=false,
            eprint=true,
            sorting=nyt,
            giveninits=true,
            maxnames=2,
            backend=bibtex]{biblatex}
\bibliography{bibliography.bib}

\setlist[itemize]{leftmargin=2em}

\title{\Velformer: Broken-Symmetry-Matched Equivariant Graph Transformers for Cosmological Velocity Reconstruction}

\author{%
  Tilman Tröster\thanks{\texttt{tilmant@phys.ethz.ch}}
  \quad
  David Mirkovic 
  \quad
  Veronika Oehl 
  \quad
  Arne Thomsen \\
  \\
  Department of Physics\\
  ETH Zürich \\
}

\begin{document}

\maketitle

\begin{abstract}
Precise measurement of the kinematic Sunyaev-Zel'dovich (kSZ) effect -- a probe of the large-scale distribution of baryonic matter, a key observable for cosmological inference -- requires accurate reconstruction of galaxy velocities from spectroscopic surveys. The signal-to-noise ratio (SNR) of kSZ measurements scales directly with the correlation coefficient $r$ between reconstructed and true velocities.

We introduce \Velformer, an equivariant graph transformer architecture designed to match the specific symmetry of the observational data. While the underlying physics is equivariant with respect to translations and rotations, observational effects break this symmetry due to the preferred line-of-sight direction. Matching the model's inductive bias to the data's broken symmetry consistently improves performance across all model sizes and training volumes, with \Velformer improving $r$ by 35\% over the standard linear theory baseline and outperforming ML baselines at every data volume.

By matching the model's inductive bias to the data and conditioning on the physics-based long-wavelength solution, \Velformer is highly data-efficient, training to high accuracy on as few as 4 low-fidelity simulations, and generalises zero-shot across input geometry, cosmological parameters, and galaxy sample. On high-fidelity simulated galaxy catalogues, this yields a 30\% improvement in $r$ over the physical baseline, directly translating to the same SNR gain on observational data.

\end{abstract}

\section{Introduction}
Contemporary and upcoming cosmological analyses are hampered not by our understanding of dark matter but by our limited knowledge of the distribution of ordinary, baryonic matter \citep{Chisari2019}. 
The kinematic Sunyaev-Zel'dovich (kSZ) effect \citep{Sunyaev1980-ksz} -- a secondary cosmic microwave background (CMB) anisotropy caused by inverse Compton scattering due to bulk velocity of free electrons -- provides a clean probe of the distribution of gas around galaxies and clusters, and is therefore a key observable for baryonic matter science.
Measuring the kSZ effect around galaxies requires knowledge about their intrinsic, or peculiar, velocities \citep{Shao2011-ksz}.

While spectroscopic galaxy surveys, such as BOSS \citep{Dawson2013-BOSS} and DESI \citep{DESI-Collaboration2016a}, map out the three-dimensional distribution of galaxies, they provide a static view of the large-scale structure of the Universe and cannot measure the velocity field directly. 
Reconstructing velocities from the observed galaxy positions can be framed as a point cloud regression problem: given galaxy 3D positions as input, predict the 3D velocity vector per point.
Inferring velocities from the galaxy distribution is a challenging inverse problem however, due to the non-linear growth of structure and confounding observational effects. 
Spectroscopic surveys measure galaxy spectra and their redshifts, from which their distance can be inferred given a model of the cosmic expansion history. 
Peculiar velocities along the line of sight (LOS) induce a Doppler shift of the galaxy spectra that are indistinguishable from the cosmic redshift however, causing a displacement of the inferred positions: the so-called redshift-space distortions (RSD).

Current approaches rely on the linearised continuity equation to estimate the velocity field from the observed galaxy density.
This linear estimate is accurate on large scales but fails to capture small-scale non-linear dynamics. 
This reliance on linear velocity reconstruction directly affects the detection significance of the kSZ effect: the signal-to-noise ratio (SNR) of kSZ measurements is proportional to the correlation coefficient between the true and inferred velocities along the LOS \citep{Ried-Guachalla2024-velocity-recon}. 
Any improvement in the velocity reconstruction thus directly translates into an increase in the detection significance of the kSZ effect. 
As the kSZ has a much lower SNR than the primary CMB and most of its secondary anisotropies, such as the thermal Sunyaev-Zel'dovich effect, improving velocity reconstruction is key to enabling downstream kSZ science.

We introduce \Velformer, an equivariant graph transformer based on Equiformer~V2 \citep{Liao2024-equiformerv2}. 
While the underlying gravitational physics is equivariant under the Euclidean group $E(3)$ of translations and rotations, the observed data is not:
the aforementioned RSDs induce a preferred direction along the LOS. 
Furthermore, our observations lie on the light-cone: galaxies further away were observed at an earlier cosmic time. 
The LOS direction thus encodes both spatial and temporal information. 
We account for this broken symmetry in \Velformer by embedding the LOS coordinate as a scalar feature, and find that this consistently improves performance across all model sizes and training-data volumes, confirming earlier findings that matching the symmetry of the model to that of the data improves performance and data efficiency \citep[e.g.][]{Fuchs2020-se3-transformer}.

We factorise the velocity reconstruction problem into large and small scales by first solving for the long-wavelength velocity field using the standard linear-theory approach and then conditioning \Velformer on this linear velocity estimate while training on small sub-volumes of size ${\sim}70\hMpc$. 
This increases the number of training samples by a factor of $\mathcal{O}(10^3)$, enabling training on as few as four simulation boxes. \Velformer achieves a $35\,\%$ improvement in $r$ over the linear theory baseline with 38 training boxes and outperforms all ML baselines. 
The model generalises zero-shot to unseen input geometries, cosmological parameters, and galaxy samples with $\gtrsim30\,\%$ improvements throughout. Achieving an equivalent gain in kSZ SNR by enlarging the galaxy survey would require approximately $70$--$80\,\%$ more galaxies.

\paragraph{Contributions}
\begin{itemize}
    \item We introduce \Velformer, an equivariant graph transformer matched to the broken $E(3)$ symmetry of spectroscopic survey data, adapted from Equiformer~V2 \citep{Liao2024-equiformerv2}.
    \item We factorise the velocity reconstruction problem into large and small scales by conditioning on the linear velocity estimate and training on small sub-volumes, increasing the number of training samples by $\mathcal{O}(10^3)$ and enabling training on as few as four simulation boxes.
    \item \Velformer outperforms all baselines at every data volume and model size, and generalises zero-shot across input geometries, cosmological parameters, and galaxy samples.
\end{itemize}

\section{Related work}
\label{sec:related}

\paragraph{ML-based velocity reconstruction}
Classical peculiar velocity reconstruction relies on linearised solutions to the continuity equation relating the density and velocity fields~\citep{Nusser1991,Zaroubi1995}. 
Deep learning approaches have recently surpassed these linear baselines by learning the non-linear density-velocity mapping directly from simulations. 
Most of these works adopt CNN or U-Net \citep{Ronneberger2015-unet} architectures operating on voxelised density grids~\citep{Tanimura2022-cnn-vel-recon,Ganeshaiah-Veena2023-unet-lpt-vel-recon,Wu2023-unet-vel-recon,Wang2024-ml-vel-recon,Xiao2025-ml-vel-recon}, while \citet{Chen2024-mlp-vel-recon} demonstrate competitive performance with a simpler MLP. 
\citet{Maragliano2026-ml-density-recon} address the related task of real-space density reconstruction rather than velocity, while \citet{Tanimura2024-graph-vel-recon} replace the grid with a GNN acting on sparse galaxy catalogues. 
The \CosmoBench \citep{Huang2025-cosmobench} dataset also provides a GNN baseline for the velocity reconstruction task. 
Our work differs from all of the above by both operating directly on halo point clouds and building in the physical symmetries of the problem by construction.

\paragraph{Equivariant neural networks}
Equivariant networks encode the requirement that outputs transform according to the symmetry group of the data. 
\citet{Cohen2016} introduced group-equivariant CNNs, followed by a general theoretical treatment in \citet{Kondor2018}. Continuous SE(3) equivariance for 3D point clouds was realised by tensor field networks~\citep{Thomas2018} and later SE(3)-Transformers~\citep{Fuchs2020-se3-transformer}, while SEGNNs~\citep{Brandstetter2022-segnn} generalise message passing to steerable node and edge features. 
Equiformer~\citep{Liao2023-equiformer,Liao2024-equiformerv2} scales these ideas with transformer-style attention. 
These architectures have found strong adoption in molecular dynamics and particle physics, where Euclidean symmetries are exact.

\paragraph{Equivariant neural networks in astrophysics}
Much existing work on symmetry-aware learning in astrophysics has targeted signals on the sphere, such as spherical CNNs~\citep{Perraudin2019,Defferrard2020-deepsphere} and wavelet-based scattering transforms~\citep{McEwen2022}. 
For 3D catalogue data, \citet{Thiele2022} apply rotationally equivariant DeepSets \citep{Zaheer2017-deepsets} to predict the thermal Sunyaev-Zel'dovich field from dark matter particle sets. \citet{Dai2022} enforce translation and rotation equivariance in a normalizing flow for field-level cosmological inference. 
More recently, \citet{Jagvaral2025} use geometric deep learning on halo graphs to model galaxy intrinsic alignments. 
We build on this tradition but target vector-valued outputs that are required to be $SE(3)$-equivariant, rather than merely invariant, and address the fact that observational effects break the $E(3)$ symmetry.

\section{Method}
\label{sec:method}

\subsection{Task definition}
\label{sec:task}

The velocity reconstruction task can be cast as a point cloud regression problem, where the 3D galaxy positions serve as inputs and the targets are the per-galaxy 3D velocity vectors. 
We refine this task by conditioning the regression on the long-wavelength velocity field obtained from linear theory, as detailed in Sects.~\ref{sec:data} and \ref{sec:vel_recon}. 
In practice this is done by including the per-galaxy linear velocity estimate as a feature vector for each input point. 
That is, we wish to find $f: \mathbb{R}^{3\times\ngal}\times \mathbb{R}^{3\times\ngal}\rightarrow \mathbb{R}^{3\times\ngal}$ that predicts velocities for all \ngal galaxies in a sample:
\begin{equation}
    f\left(\{\vec{x}_i, \vlin_i\}_{i=1}^\ngal\right) \rightarrow \{\vpred_i\}_{i=1}^\ngal\,,
\end{equation}
where $\vec{x}_i$, $\vlin_i$, and $\vpred_i$ are the 3D galaxy positions, linear velocity estimates, and predicted velocities for each galaxy, respectively. 

The optimisation proceeds by minimising the mean-squared-error (MSE) loss between \vpred and the true velocities \vtrue, see Eq.~\ref{equ:metrics}. 
Each training sample is given in the form of a collection of \ngal galaxies $D_{i} = \{\vec{x}_{i}, \vlin_{i}, \vtrue_{i}\}$, $i=1,\dots,\ngal$, where \ngal can vary from sample to sample. 

For the graph-based models, the point cloud is transformed into a graph with node features $\vec{x}_i$ and \vlin, and edge features comprised of the relative positions $\vec{x}_i-\vec{x}_j$ and distance $|\vec{x}_i-\vec{x}_j|$. 
Graph connectivity is defined by a $k$ nearest neighbours (\knn) search, as described in Sect.~\ref{sec:graph-construction}.

\subsection{\Velformer}
\Velformer is an equivariant graph Transformer architecture, closely following that of Equiformer V2 \citep{Liao2024-equiformerv2}. 
It inherits Equiformer V2's features, such as efficient eSCN \citep{Passaro2023-eSCN} convolutions that reduce the $SO(3)$ tensor products to  $SO(2)$ linear operations. 
We refer the reader to \citet{Liao2024-equiformerv2} for architectural details about Equiformer V2 and limit ourselves here to changes relevant for \Velformer.

The input linear velocity estimates $\vlin_i$ are embedded as spherical harmonics up to degree \lmax, scaled by the norm of the velocity vectors. 
The output \vpred are taken as the final vector ($L{=}1$) irreducible representations (irreps). 
In order to break the $E(3)$ equivariance, we embed the LOS coordinate as a scalar, i.e. $L{=}0$ irrep. 
This procedure is similar to how SE(3)-Transformer+z \citep{Fuchs2020-se3-transformer} accounts for the preferred direction induced by gravity in real-world object classification.  

In addition to the broken-$E(3)$ \Velformer, which is our fiducial model, we also consider a variant with full $E(3)$ equivariance, as well as a non-equivariant version. 
For the latter $\lmax{=}0$ and velocity in- and outputs are treated as independent scalar channels.

As a production model for non-linear velocity reconstruction will likely use application-specific training data different from ours (cf. Sect.~\ref{sec:data}), we are here primarily interested in exploring the performance of \Velformer across different data regimes and model sizes, rather than producing a single `best' model. 
To this end we consider a wide range of model sizes, ranging from approximately 50k to 200M parameters. 
The hyperparameters, model configuration, and training details are described in Appendix~\ref{app:velocityformer}.

\subsection{Baselines}
\label{sec:baselines}
We consider four baselines to compare against our proposed \Velformer architecture: the standard linear-theory velocity reconstruction, as described in Appendix~\ref{sec:vel_recon}; a U-Net architecture, to benchmark against previous ML approaches to velocity reconstruction; the GNN baseline introduced with the \CosmoBench \citep{Huang2025-cosmobench} dataset; and a standard Transformer architecture, as an architecture with no geometric inductive bias. 
We also investigated the Geometric Algebra Transformer \citep{Brehmer2023-geometric-algebra-transformer} as an alternative equivariant architecture but found training to be unstable and it failing to outperform the simpler baselines.

Architectural details and training hyperparameters are described in Appendix~\ref{app:training}.

\subsubsection{Physical baseline: linear theory}
\label{sec:linear_vel_recon_baseline}
The physical baseline is given by the linear-theory velocity reconstruction described in Appendix~\ref{sec:vel_recon}, the current de facto standard tool in kSZ analyses. 
It therefore serves as a key benchmark for our proposed method, as the main motivation for this work is to improve upon the linear estimator and thus increase the SNR of kSZ measurements. 
As the ML-based models are given the linear velocity estimate as an input, this baseline also serves as a floor for the ML model performance, since learning the identity mapping recovers the linear estimate. 

\subsubsection{U-Net baseline}
\label{sec:unet_baseline}
Much of prior work \citep[e.g.][]{Ganeshaiah-Veena2023-unet-lpt-vel-recon,Wu2023-unet-vel-recon,Wang2024-ml-vel-recon,Xiao2025-ml-vel-recon} on ML-based velocity reconstruction has focused on convolutional neural network architectures, in particular U-Nets \citep{Ronneberger2015-unet}. 
To compare to these previous works, we implement a similar U-Net baseline. 
Unfortunately, there is no established benchmark dataset, public implementation, or even agreed-upon metric among these prior works. 
While we find similar results with our U-Net implementation and training regime, making a direct comparison remains difficult. 

The input point cloud (galaxy catalogue) is first converted to a voxelised density field on a regular grid using a cloud-in-cell mass-assignment scheme, which is fed into the U-Net. 
The predicted velocity grid is then linearly interpolated at the galaxy positions to obtain per-galaxy velocity estimates.

\subsubsection{GNN baseline}
\label{sec:gnn_baseline}
\CosmoBench \citep{Huang2025-cosmobench} introduced a benchmark dataset based on, among others, the same \Quijote simulations we use in this work (see Sect.~\ref{sec:data}), and proposed a GNN architecture for the velocity reconstruction task. 
While the dataset definition differs from ours (e.g. they only consider the 5000 most massive haloes in each box), the GNN architecture serves as a useful point of comparison for our proposed \Velformer, as it operates on the same graph structure as our model, but without the inductive bias of equivariance.
We slightly adapt their architecture to our problem setup, in particular by conditioning the model on the linear velocity estimate.

\subsubsection{Transformer baseline}
\label{sec:transformer_baseline}
While the other baselines include some form of inductive bias (translational equivariance due to the convolutional structure of the U-Net, locality due to the \knn graph of the GNN), the Transformer architecture operates on all galaxies in the sub-boxes (see Sect.~\ref{sec:data}) as a set, with no built-in assumptions about the underlying symmetries or structure of the data.

The self-attention layers do not use positional encodings, leaving the model to learn the geometric structure from the galaxy positions. 
The fiducial Transformer baseline attends to all galaxies in the input but we test using the graph adjacency matrix as an attention mask to emulate the behaviour of the graph-based models in Appendix~\ref{app:varyk}.

\section{Experimental setup}
\label{sec:experiments}
 
\subsection{Data}
\label{sec:data}

The training and evaluation data for our experiments are based on the \Quijote \citep{Villaescusa-Navarro2020-quijote} $N$-body simulation suite. 
While the resolution of the \Quijote simulations is relatively low compared to state-of-the-art cosmological simulations, the very large number of realisations makes them an ideal test bed for ML applications.

We use the `fiducial' set, which has a fixed cosmology but varies in the initial conditions. 
The simulation snapshots are in the form of cubic boxes with side-length of $1\,h^{-1}\mathrm{Gpc}$ and periodic boundary conditions. 
To produce the mock galaxy samples used for training, validation, and testing, we start with the halo catalogues from the simulation snapshots at redshift $z{=}0.5$. 
The haloes are populated with galaxies with a halo-occupation distribution (HOD) model, following \citet{Zheng2007} and using the implementation provided by \texttt{pyHOD}.\footnote{\url{https://gitlab.com/jcallesh/hod}}  
Velocities of central galaxies are set to the velocity of their host halo, while satellite galaxies are assigned a random velocity component drawn from a Gaussian distribution with variance given by the virial velocity of the host halo. 
The use of random satellite velocities makes the reconstruction task arguably more challenging than on observational data, as it adds a stochastic component to the velocity field that is not derived from gravitational dynamics. 
Finally, the galaxy catalogues are converted to redshift-space by applying the line-of-sight displacement Eq.~\ref{equ:rsd}. 
The resulting mock galaxy catalogues emulate the BOSS CMASS sample, with a number density of $n \approx 3.5\times 10^{-4}\ihMpcV$ and a linear bias of $b \approx 2$.

Linear velocity estimates are then obtained for each full $1 \hGpcV$ simulation box using a standard FFT-based estimator for Eq.~\ref{eq:v_lin_rsd}, with a grid size of $256^3$ and a Gaussian smoothing kernel with a width of $10 \hMpc$.

Each galaxy catalogue constitutes a point cloud: an unordered set of 3D positions with associated feature vectors, the linear velocity estimates, and per-point true velocities as regression targets. 
Sample galaxy point clouds are visualised in Fig.~\ref{fig:sample_graphs}. 

In this work we consider three data regimes:
\begin{itemize}
    \item \textbf{Low-data}: 4 simulation boxes for training, corresponding to $\sim 10^4$ sub-boxes, i.e. training samples. This regime is a realistic scenario for training on high-fidelity hydrodynamic simulations, for example \flamingo \citep{Schaye2023-flamingo, Kugel2023-flamingo, Helly2026-flamingo-data-release}, 
    which are computationally expensive and thus limited in number, especially for cosmology-sized boxes.
    \item \textbf{Mid-data}: 38 simulation boxes for training, corresponding to $\sim 10^5$ sub-boxes. This data volume is consistent with high-resolution $N$-body simulation suites, such as the Aemulus Project \citep{DeRose2019}. 
    \item \textbf{High-data}: 3800 simulation boxes for training, corresponding to $\sim 10^7$ sub-boxes. While this data volume is not available for high-fidelity simulations we would use for production training runs, it allows us to explore the large-data regime and confirm that the equivariant model continues to outperform baselines even when data is abundant.
\end{itemize}

Unless noted otherwise, results are reported on a held-out test set of 50 simulation boxes, corresponding to $\sim 1.3\times 10^5$ sub-boxes. 
The uncertainty estimation procedure is described in Appendix~\ref{app:errors}.

\subsection{Factorisation into large and small scales}
As the long-wavelength velocity field is captured by the linear estimator, we factorise the problem into large and small scales by subdividing the $1\hGpcV$ simulation boxes into $\nsplit^3$ sub-boxes. 
The models are then trained to predict the non-linear velocity field on the sub-boxes, conditioned on the linear velocity estimated from the full box. 
We find that $\nsplit{=}14$ provides a good balance between capturing non-linear scales and having enough training samples, resulting in sub-boxes of size $71.4\hMpc$.
In Appendix~\ref{app:varynsplit} the effect of the choice of \nsplit is investigated in detail.

This factorisation brings three main benefits:
\begin{itemize}
      \item It grounds the model in the large-scale velocity field, which is based on well-understood physics and is accurately captured by the linear estimator.
      \item The sub-boxes contain $\mathcal{O}(100)$ galaxies, making the use of graph-based models, which would struggle with $\mathcal{O}(10^5)$ nodes and $\mathcal{O}(10^6)$ edges contained in the full box, computationally feasible.
      \item It allows us to increase the number of training samples by a factor of $\mathcal{O}(10^3)$, thus enabling robust training even in data-scarce scenarios with $\mathcal{O}(1)$ simulations.
\end{itemize}

\subsection{Graph construction}
\label{sec:graph-construction}
Both our proposed \Velformer architecture and the GNN baseline operate on graphs. 
We thus interpret the galaxy catalogues as graphs, with node features being given by the three-dimensional galaxy positions and linear velocity predictions. 
Edges are defined using a $k$-nearest neighbour (\knn) graph built from the redshift-space positions of the galaxies in each sub-box. 
We choose \knn graphs over radius graphs, as they ensure connectivity of the graph even in low-density regions, avoiding disconnected clusters of nodes. 
 
We find that using a graph structure with $k{=}10$ nearest neighbours strikes a good balance between capturing local interactions and keeping the computational cost manageable. 
Imposing such a graph structure is an arbitrary choice, without physical motivation, as gravity is a long-range force that induces interactions between all galaxies, regardless of their distance.
Indeed, we find that increasing the connectivity of the graph does improve model performance, but at increasingly prohibitive computational cost, even for small models. 
Details on the effect of varying the connectivity of the graph on model performance are given in Appendix~\ref{app:varyk}.

\subsection{Evaluation metrics}
\label{sec:metrics}

We consider two main metrics to evaluate model performance: the mean-squared error (MSE) $l$ of the predicted three-dimensional velocities and their ground truths; and the correlation coefficient $r$ between the predicted and true velocities along the LOS. 
For predicted velocities $v^\mathrm{pred}_{ij}$, $i=1,\dots,N$, $j=1,2,3$, and true velocities $v^\mathrm{true}_{ij}$, 
\begin{equation}
\label{equ:metrics}
      l = \frac{1}{3N} \sum_{i=1}^N \sum_{j=1}^3 (v^\mathrm{pred}_{ij} - v^\mathrm{true}_{ij})^2\,, \quad r = \frac{1}{\sigma_{v^\mathrm{pred}}\sigma_{v^\mathrm{true}}} \frac{1}{N} \sum_{i=1}^N v^\mathrm{pred}_{i,\parallel} v^\mathrm{true}_{i,\parallel}\,,
\end{equation}
where $\sigma_{v^\mathrm{pred/true}}$ are the standard deviations of the predicted and true LOS velocities, respectively, and $v_{i,\parallel}$ is the LOS velocity of galaxy $i$. 
In our case, the LOS is aligned with the $z$-axis of the simulation box, so $v_{i,\parallel} = v_{i,3}$. 

The MSE $l$ captures the overall accuracy of the velocity predictions, and serves as the training loss for all models, while the LOS correlation coefficient $r$ is directly related to the SNR of kSZ measurements and thus serves as a key metric for our main scientific application. 
We furthermore quote \Deltar, the relative improvement in $r$ over the physical baseline, as an estimate of the improvement in kSZ SNR that our method would enable.

\section{Results}
\label{sec:results}

Our main results for the cases of 4 and 38 simulation boxes are shown in Table~\ref{tab:best-l-r-results}. 
The quoted numbers are for the model size with the best MSE; see Table~\ref{tab:size_comparison} in Appendix~\ref{app:extended-results} for the results at each model size. 
The \Velformer architecture consistently outperforms all baselines across all model sizes and training-data volumes, confirming the importance of matching the symmetry of the model to that of the data. 
It reaches a 30\% improvement over the physical baseline with just 4 training boxes, and 35\% on 38 boxes, offering significant gains in the SNR of kSZ measurements. 
Achieving a comparable gain in SNR through increasing the number of galaxies the kSZ effect is measured on would require a 70--80\% larger galaxy catalogue.

In the regimes with limited training data, increasing model size for the broken-symmetry \Velformer has a limited effect, with performance staying within few per cents over model size increases by three orders of magnitude. 
In contrast, the full-$E(3)$ variant benefits from larger model capacity, particularly in the data-sparse 4-boxes regime, suggesting that an improper inductive bias can be compensated with more model capacity and more data.

\begin{table}[ht]
  \centering
  \caption{%
    Velocity MSE $l$, LOS velocity correlation coefficient $r$ (see Eq.~\ref{equ:metrics}), and improvement over the physical (linear theory) baseline $\Delta r$ for each model architecture and two data regimes: low-data (4 simulation boxes) and mid-data (38 simulation boxes). 
    For each model the parameter count that yields the lowest validation loss is selected.
    Uncertainties are $< 1\%$ for $l$ and $r$ and are not shown here for brevity, see Appendix~\ref{app:errors} for details. The uncertainties in $\Delta r$ are somewhat larger due to being a ratio of two noisy quantities.%
  }
  \label{tab:best-l-r-results}
  \small
  \begin{tabularx}{\textwidth}{lXrrrXrrr}
    \toprule
    Model & & \multicolumn{3}{c}{Low-data} & & \multicolumn{3}{c}{Mid-data} \\
    \cmidrule(lr){3-5} \cmidrule(lr){7-9}
    & & $\downarrow l$ & $\uparrow r$ & \multicolumn{1}{l}{\hspace{6pt}$\uparrow \Delta r$ [\%]} & & $\downarrow l$ & $\uparrow r$ & \multicolumn{1}{l}{\hspace{6pt}$\uparrow \Delta r$ [\%]} \\
    \midrule
    Physical baseline & & 0.825 & 0.571 & -- & & 0.825 & 0.571 & -- \\
    \midrule
    Transformer & & 0.651 & 0.721 & $25.7 \pm 1.3$ & & 0.614 & 0.753 & $31.3 \pm 1.3$ \\
    GNN & & 0.671 & 0.718 & $25.6 \pm 0.2$ & & 0.639 & 0.738 & $28.8 \pm 0.2$ \\
    U-Net & & 0.677 & 0.708 & $20.8 \pm 0.2$ & & 0.652 & 0.734 & $25.5 \pm 0.2$ \\
    \midrule
    Velocityformer $L_\mathrm{max}{=}0$ & & 0.733 & 0.643 & $12.4 \pm 0.4$ & & 0.677 & 0.688 & $21.2 \pm 0.5$ \\
    Velocityformer $E(3)$ & & 0.652 & 0.728 & $26.8 \pm 0.3$ & & 0.600 & 0.765 & $35.1 \pm 0.3$ \\
    Velocityformer broken-$E(3)$ & & \textbf{0.627} & \textbf{0.741} & $\mathbf{29.9 \pm 0.4}$ & & \textbf{0.596} & \textbf{0.779} & $\mathbf{35.9 \pm 0.4}$ \\
    \bottomrule
  \end{tabularx}
\end{table}

\subsection{Data efficiency}
\label{sec:data_efficiency}

\begin{figure}
      \centering
      \includegraphics[width=\textwidth]{"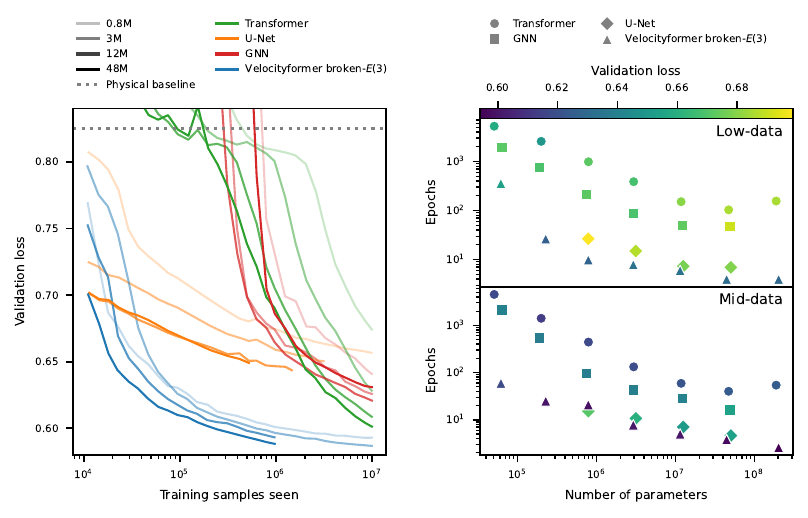"}
      \vspace{-2em}
      \caption{\emph{Left}: validation loss curves over a single epoch of for the high-data regime (3800 simulation boxes). \Velformer (blue lines) learns faster than the baselines, reaching a fixed loss with 10--100 times fewer training samples. \emph{Right}: number of epochs required for convergence for different model sizes. The top panel shows training on 4 simulation boxes (low-data), while the case for 38 simulation boxes (mid-data) is shown on the bottom panel. \Velformer (triangle symbols) converges faster and to a lower loss than the baselines.}
      \label{fig:params-vs-epochs}
\end{figure}

An inductive bias that better matches the structure of the data generally improves data efficiency, both in terms of number of epochs required for convergence and in terms of the number of training samples required to reach a given performance level. 
This is illustrated in the right panel of Fig.~\ref{fig:params-vs-epochs}, which lists the number of epochs required for each model to reach convergence. 
\Velformer converges in 10--100 times fewer epochs than the Transformer and GNN baselines while achieving a lower loss. 
The U-Net baseline requires a similar number of epochs for convergence as \Velformer, potentially due to its partially equivariant nature, but does not achieve the same low loss.

The left panel of Fig.~\ref{fig:params-vs-epochs} shows the validation loss curves for \Velformer and the baselines for the data-abundant case of 3800 training boxes. 
\Velformer reaches any given performance level with a factor of 10--100 less training data than the baselines, again demonstrating its data efficiency.  

Heuristically, \Velformer exhibits a steeper exponent compared to the baselines when fitting a power law with finite offset of the form $L(d) = L_\mathrm{min} + \frac{A}{d^\alpha}$ to the loss-vs-data curve, as in \citet{Kaplan2020-llm-scaling, Hoffmann2022-chinchilla, ATLAS2026-scaling, Vigl2026-jet-transformer-scaling-law}. 
We found the power law fit to be highly sensitive to initialisation and fitting range, however, and thus refrain from making quantitative statements here but encourage further work in this direction.

\subsection{Zero-shot generalisation}
\label{sec:zero-shot}
The training data used in our experiments is based on a single cosmology and HOD prescription. 
This is a somewhat unrealistic scenario, as any production model would be trained on a more diverse set of data, both in terms of cosmological parameters and galaxy sample. 
Nonetheless, we investigate the zero-shot generalisation of \Velformer as a stress test of the model's capability to bridge the sim-to-real gap. 

We consider three test cases, with an increasing distribution drift from the training data: using a different \nsplit at test time compared to training; an unseen simulation with a different cosmology from the \texttt{nwLH} latin hypercube set of the \Quijote suite \citep{Villaescusa-Navarro2020-quijote}; and a high-fidelity galaxy catalogue from the \flamingo simulation suite \citep{Schaye2023-flamingo, Kugel2023-flamingo, Helly2026-flamingo-data-release}. 
The first test probes the model's ability to generalise to different input geometries, a key capability for real-world applications where the data will not fit neatly into sub-boxes of a fixed size. 
The second test assesses the model's robustness to changes in cosmological parameters, as the true cosmology of the Universe is unknown. 
Indeed, the present work is largely in service to ultimately enabling better inference of cosmological parameters. 
The third test evaluates the model's capacity to generalise from the crude HOD-based galaxy catalogues used for training to a more realistic galaxy catalogue that is based on a full hydrodynamical simulation and galaxy formation model. 
Details of these different datasets are described in Appendix~\ref{app:zero-shot-setup}.

\begin{table}[ht]
  \centering
  \caption{%
    Zero-shot generalisation of \Velformer across different sub-box sizes, cosmologies, and galaxy sample.
    Uncertainties are estimated from the scatter between five simulation boxes from the test set; \flamingo uncertainties are scaled by~1.7 to account for the smaller box size.%
  }
  \label{tab:zeroshot}
  \begin{tabular}{lrr@{\hskip 10pt}rr@{\hskip 10pt}r}
    \toprule
    & \multicolumn{2}{c}{Sub-box size} & \multicolumn{2}{c}{Cosmology} & \flamingo \\
    \cmidrule(lr){2-3} \cmidrule(lr){4-5} \cmidrule(lr){6-6}
    & $83.3\,\hMpc$ & $50\,\hMpc$ & A & B \\
    \midrule
    $r_{\rm lin}$ & $0.582 \pm 0.013$ & $0.584 \pm 0.013$ & $0.565 \pm 0.013$ & $0.607 \pm 0.013$ & $0.549 \pm 0.022$ \\
    $r$ & $0.754 \pm 0.005$ & $0.763 \pm 0.005$ & $0.765 \pm 0.005$ & $0.783 \pm 0.005$ & $0.722 \pm 0.009$ \\
    $\Delta r$ [\%] & $30.0 \pm 2.1$ & $32.1 \pm 2.1$ & $36.4 \pm 2.1$ & $29.7 \pm 2.1$ & $32.6 \pm 3.6$ \\
    \bottomrule
  \end{tabular}
\end{table}

These zero-shot generalisation results are presented in Table~\ref{tab:zeroshot}, showing that a 0.8M parameter \Velformer trained on 38 simulation boxes is able to maintain a significant performance improvement over the physical baseline even in the face of substantial distribution shifts. 
This robustness makes \Velformer a strong contender for application on real observational data. 
We would like to emphasise that a production model would be trained on a more diverse set of data, capturing both variations in cosmological parameters and the galaxy sample, and thus would be expected to generalise even better than the zero-shot results presented here. 
These more realistic training scenarios are being investigated in forthcoming work.

\section{Discussion \& conclusion}
\label{sec:discussion}

The main limitation that prevents application of \Velformer at a production level is the simplified setup of the training data. 
For application to observational data, the following upgrades should be implemented on the training data to address the sim-to-real gap:
\begin{itemize}
    \item \textbf{Redshift coverage, light-cone data, and survey geometry} As observed galaxy samples cover a wide range of redshifts (the various DESI samples cover redshifts from $z{=}0.1$ to $z{=}3.5$ for quasars; \citealt{DESI-Collaboration2026-dr1}), the training data should reflect this diversity. Furthermore, the model should be confirmed to generalise to light-cone data on irregular survey geometries, rather than the regular boxes at fixed redshift used here. 
    \item \textbf{Cosmology and galaxy sample} The true cosmological parameters of the Universe are not precisely known and reproducing realistic galaxies in simulation is still a very challenging task. 
    The training data should reflect these uncertainties in our knowledge of cosmological parameters and galaxy formation.  
    \item \textbf{Regression target} For kSZ measurements, the true quantity of interest is the bulk velocity of the gas surrounding the galaxies. Here we use the galaxy velocities as the regression target, a proxy for gas bulk velocity. This can be refined to regressing on the velocities of the haloes that host the galaxies, which correlates better with the gas velocity \citep{Ried-Guachalla2024-velocity-recon}, or regressing directly on the gas velocity using high-fidelity hydrodynamical simulations.
\end{itemize}

The design choices and data efficiency of \Velformer make it feasible to address these limitations. 
The factorisation into large and small scales employed by \Velformer is an important asset, as light-cone effects are negligible on the scale of the small sub-boxes. 
By operating on small point clouds, rather than fixed regular grids, such as the U-Net approach, \Velformer can also easily adapt to irregular survey geometries, as the sub-box point clouds are generally smaller than features in the survey geometry. 

The data efficiency -- being able to improve over the physical baseline by 30\% with only 4 simulation boxes -- would in principle allow \Velformer to be trained directly on light-cone data from high-fidelity hydrodynamical simulations. 
A more promising approach might be to pre-train on the large data volumes provided by the \Quijote simulation suite, including both variations in cosmology and HOD prescription. 
The pre-trained model can then be fine-tuned on scarce high-fidelity data. 
We leave the investigation of such transfer-learning approaches for future work.

Training and inference time per sample is significantly higher on the equivariant models, especially compared to highly-optimised self-attention implementations of the Transformer. 
To a large degree this reflects engineering investments, rather than computational complexity, however. 
Ongoing efforts to optimise equivariant operations, such as {\sc{cuEquivariance}},\footnote{\url{https://github.com/NVIDIA/cuEquivariance}} will likely shrink this gap in the future.

\paragraph{Conclusions}
We introduce \Velformer, an equivariant graph transformer architecture that supports the specific broken $E(3)$ symmetry of spectroscopic galaxy survey data. 
\Velformer outperforms the baselines at all data volumes and model sizes, reaching a 35\% improvement in the LOS velocity correlation coefficient $r$  over the physical linear theory baseline, which is the current standard tool employed for velocity reconstruction. 
This gain in $r$ directly translates into a proportional gain in the SNR of kSZ measurements, improving the detection significance of this crucial probe of diffuse gas in the Universe.

The performance advantage persists over three orders of magnitude in data volumes, ranging from 4 to 3800 simulation boxes, corresponding to $10^4$--$10^7$ training samples, as well as model sizes, ranging from 50k to 200M parameters. 
\Velformer is highly data efficient, training robustly on as few as 4 simulation boxes, and generalises well to input geometries, cosmologies, and galaxy samples, still achieving $\gtrsim 30\%$ improvements over the physical baseline on these unseen examples.
This strong generalisation capability allows \Velformer to bridge the sim-to-real gap, with refinements on the training regime to further increase robustness to distribution shift being the subject of forthcoming work.
This will allow \Velformer to be confidently applied to observational data, boosting the detection significance of kSZ measurements by over 30\%.

\begin{ack}
TT and VO acknowledge funding from the Swiss National Science Foundation under the Ambizione project PZ00P2\_193352. 

We acknowledge the Virgo Consortium for making their simulation data available. The FLAMINGO simulations were performed using the Durham Memory Intensive system managed by the Institute for Computational Cosmology on behalf of the STFC DiRAC facility (www.dirac.ac.uk).
\end{ack}

\section*{References}
\medskip

\printbibliography[heading=none]

\appendix

\section{Scientific background}
\label{sec:background}

\subsection{Spectroscopic galaxy surveys, redshift-space distortions, and symmetries}

\label{app:rsd}

The standard model of cosmology rests on the \emph{cosmological principle}: the Universe is statistically homogeneous and isotropic on large scales.
Ignoring relativistic corrections, homogeneity and isotropy together imply equivariance under the Euclidean group $E(3)$ of translations and rotations. 
This symmetry is the physical motivation for equivariant architectures in large-scale structure modelling.

Spectroscopic galaxy surveys such as BOSS \citep{Dawson2013-BOSS} and DESI
\citep{DESI-Collaboration2016a} measure the angular positions of galaxies on the sky
and their redshifts, from which comoving distances are inferred given a model of the
cosmic expansion history. 
Peculiar velocities along the line of sight (LOS) induce a Doppler shift
indistinguishable from the cosmological redshift, displacing the inferred position of
each galaxy along the LOS by
\begin{equation}
  \label{equ:rsd}
  \vec{s} = \vec{x} + \frac{\vec{v}\cdot\nhat}{aH}\,\nhat\,,
\end{equation}
where $\vec{x}$ and $\vec{s}$ are the real- and redshift-space positions,
$\vec{v}\cdot\nhat$ is the LOS peculiar velocity, $a$ is the scale factor, and $H$ is
the Hubble parameter.
On large scales, coherent infall of galaxies into overdensities compresses structures
along the LOS -- the \emph{Kaiser effect} \citep{Kaiser1987a}.
On small scales, the virial motions of galaxies within clusters stretch structures into
elongated filaments along the LOS -- the so-called \emph{fingers of God}
\citep{Jackson1972-fog}.

The displacement in Eq.~\ref{equ:rsd} singles out the LOS as a preferred direction, breaking the $E(3)$ symmetry
of the underlying physics.
A second, independent source of symmetry breaking arises from the light-cone geometry of our observations: galaxies at greater distances were observed at earlier cosmic times, the LOS direction therefore encodes temporal as well as spatial information. 
Together, these observational effects reduce the effective symmetry of the data from $E(3)$ to a residual $E(2)$ in the plane of the sky, and motivate the broken-$E(3)$ design of \Velformer (Sect.~\ref{sec:method}).

\subsection{Linear velocity reconstruction}
\label{sec:vel_recon}

To first order, galaxy velocities $\mathbf{v}$ can be obtained from the linearised continuity equation. 
The continuity equation in the cosmology context is given by 
\begin{equation} 
\label{eq:continuity}
    \nabla\cdot{\mathbf{v}}=-aHf\frac{\delta_g}{b}\,,
\end{equation}
where $a$ is again the scale factor, $H$ the Hubble parameter, $f = \frac{\mathrm{d}\ln{D}}{\mathrm{d}\ln{a}}$ the linear growth rate with $D$ the growth factor, and $b$ denotes the linear bias connecting the matter density contrast $\delta_m$ to the galaxy density contrast $\delta_g = b \delta_m$. 
Adding the effect of RSDs introduces an extra term to Eq.~\ref{eq:continuity} \citep{Nusser1991, Yahil1991, Nusser1994}:
\begin{equation} 
\label{eq:continuity_rsd}
    \nabla\cdot{\vec{v}}+\frac{f}{b}\nabla\cdot[(\vec{v}\cdot\nhat)\nhat]=-aHf\frac{\delta_g}{b},
\end{equation} 
where $\nhat$ is the LOS direction. 
Inverting Eq.~\ref{eq:continuity_rsd} is not possible without specification of boundary conditions. 
Long-range modes can be recovered by linearising the equation however \citep{Nusser1991, Dekel1993, Nusser1994}.  
In Fourier space, the linearised continuity equation can be solved as \citep{Ried-Guachalla2024-velocity-recon}
\begin{equation}
\label{eq:v_lin_rsd}
  \vec{v}(\vec k) = - i a H f \frac{\vec{k}}{k^2} \frac{\delta_g(\vec k)}{(b + f\mu^2)} W(k, r_\mathrm{smooth}).
\end{equation}
Here, $k = |\vec{k}|$, and $\mu$ is the cosine of the LOS angle with respect to the mode $\vec{k}$, $\mu  = \nhat \cdot \vec{k} / k$ and $W(k, r_\mathrm{smooth})$ is a Gaussian smoothing filter to suppress small-scale noise.

Several higher-order correction methods to improve upon the linear reconstruction exist \citep[see][for an overview]{Kitaura2012} but they tend to not lead to an improvement in the correlation coefficient $r$ on realistic mock catalogues \citep{Planck-Collaboration2016-ksz}.

\section{Additional Results}
\label{app:extended-results}
In Table~\ref{tab:size_comparison} we present the results for all model sizes and data volumes, showing the velocity MSE $l$, LOS velocity correlation coefficient $r$, and improvement over the physical baseline $\Delta r$. 
Larger models do not necessarily perform better, suggesting a bias-variance trade-off, with no clear evidence for double descent.

\begin{table}[ht]
  \centering
  \caption{%
    Velocity MSE $l$, LOS velocity correlation coefficient $r$ (see Eq.~\ref{equ:metrics}), and improvement over the physical (linear theory) baseline $\Delta r$ for each model architecture for two data regimes: low-data (4 simulation boxes) and mid-data (38 simulation boxes). 
    For each model the best results are highlighted in bold.
    Uncertainties are $< 1\%$ for $l$ and $r$ and are not shown here for brevity, see Appendix~\ref{app:errors} for details.    The uncertainties in $\Delta r$ are somewhat larger due to being a ratio of two noisy quantities.%
  }
  \label{tab:size_comparison}
  \small
  \begin{tabular}{ll@{\hspace{15pt}}rrr@{\hspace{15pt}}rrr}
    \toprule
    Model & Size & \multicolumn{3}{c}{Low-data} & \multicolumn{3}{c}{Mid-data} \\
    \cmidrule(lr){3-5} \cmidrule(lr){6-8}
    & & $\downarrow l$ & $\uparrow r$ & \multicolumn{1}{l}{\hspace{6pt}$\uparrow \Delta r$ [\%]} & $\downarrow l$ & $\uparrow r$ & \multicolumn{1}{l}{\hspace{6pt}$\uparrow \Delta r$ [\%]} \\
    \midrule
    Physical baseline & & 0.825 & 0.571 & -- & 0.825 & 0.571 & -- \\
    \midrule
    \multirow{7}{*}{Transformer} & 0.05M & 0.660 & 0.710 & $23.8 \pm 1.3$ & 0.619 & 0.749 & $30.7 \pm 1.3$ \\
     & 0.2M & \textbf{0.651} & \textbf{0.721} & $\mathbf{25.7 \pm 1.3}$ & \textbf{0.614} & \textbf{0.753} & $\mathbf{31.3 \pm 1.3}$ \\
     & 0.8M & 0.672 & 0.696 & $21.3 \pm 1.3$ & 0.619 & 0.749 & $30.6 \pm 1.3$ \\
     & 3M & 0.669 & 0.699 & $22.1 \pm 1.3$ & 0.620 & 0.748 & $30.5 \pm 1.3$ \\
     & 12M & 0.681 & 0.684 & $19.6 \pm 1.3$ & 0.624 & 0.745 & $30.1 \pm 1.3$ \\
     & 48M & 0.687 & 0.679 & $18.8 \pm 1.3$ & 0.626 & 0.744 & $30.1 \pm 1.3$ \\
     & 200M & 0.686 & 0.679 & $18.7 \pm 1.3$ & 0.626 & 0.744 & $30.1 \pm 1.3$ \\
    \midrule
    \multirow{6}{*}{GNN} & 0.05M & 0.674 & 0.714 & $24.4 \pm 0.2$ & 0.641 & 0.737 & $28.5 \pm 0.2$ \\
     & 0.2M & 0.671 & 0.716 & $24.9 \pm 0.2$ & \textbf{0.639} & 0.738 & $28.8 \pm 0.2$ \\
     & 0.8M & 0.673 & 0.716 & $25.0 \pm 0.2$ & 0.640 & 0.738 & $28.9 \pm 0.2$ \\
     & 3M & \textbf{0.671} & \textbf{0.718} & $25.6 \pm 0.2$ & 0.640 & \textbf{0.738} & $29.1 \pm 0.2$ \\
     & 12M & 0.675 & 0.716 & $\mathbf{25.6 \pm 0.2}$ & 0.642 & 0.737 & $\mathbf{29.4 \pm 0.2}$ \\
     & 48M & 0.693 & 0.705 & $23.6 \pm 0.2$ & 0.656 & 0.732 & $28.3 \pm 0.2$ \\
    \midrule
    \multirow{4}{*}{U-Net} & 0.8M & 0.698 & 0.664 & $16.7 \pm 0.2$ & 0.672 & 0.693 & $21.8 \pm 0.2$ \\
     & 3M & 0.686 & 0.675 & $19.0 \pm 0.2$ & 0.661 & 0.702 & $23.7 \pm 0.2$ \\
     & 12M & \textbf{0.677} & \textbf{0.708} & $\mathbf{20.8 \pm 0.2}$ & \textbf{0.652} & \textbf{0.734} & $\mathbf{25.5 \pm 0.2}$ \\
     & 48M & 0.678 & 0.686 & $20.6 \pm 0.2$ & 0.653 & 0.712 & $25.2 \pm 0.2$ \\
    \midrule
    \multirow{6}{*}{Velocityformer $L_\mathrm{max}{=}0$} & 0.05M & 0.739 & 0.635 & $11.1 \pm 0.4$ & 0.719 & 0.656 & $14.7 \pm 0.5$ \\
     & 0.2M & 0.735 & 0.639 & $11.8 \pm 0.4$ & 0.698 & 0.673 & $17.8 \pm 0.5$ \\
     & 0.8M & 0.738 & 0.638 & $11.5 \pm 0.4$ & 0.682 & 0.688 & $20.3 \pm 0.5$ \\
     & 3M & \textbf{0.733} & \textbf{0.643} & $12.4 \pm 0.4$ & 0.682 & 0.688 & $20.3 \pm 0.5$ \\
     & 12M & 0.738 & 0.637 & $11.5 \pm 0.4$ & 0.677 & \textbf{0.690} & $20.9 \pm 0.5$ \\
     & 48M & 0.734 & 0.640 & $\mathbf{12.6 \pm 0.5}$ & \textbf{0.677} & 0.688 & $\mathbf{21.2 \pm 0.5}$ \\
    \midrule
    \multirow{7}{*}{Velocityformer $E(3)$} & 0.05M & 0.765 & 0.599 & $4.7 \pm 0.2$ & 0.669 & 0.704 & $23.1 \pm 0.3$ \\
     & 0.2M & 0.699 & 0.669 & $17.0 \pm 0.3$ & 0.612 & 0.755 & $32.2 \pm 0.3$ \\
     & 0.8M & 0.698 & 0.670 & $17.8 \pm 0.3$ & 0.606 & 0.760 & $33.6 \pm 0.3$ \\
     & 3M & 0.688 & 0.689 & $20.6 \pm 0.3$ & 0.610 & 0.756 & $32.4 \pm 0.3$ \\
     & 12M & 0.681 & 0.691 & $21.5 \pm 0.3$ & 0.610 & 0.756 & $32.9 \pm 0.3$ \\
     & 48M & 0.657 & 0.715 & $26.1 \pm 0.3$ & \textbf{0.600} & 0.765 & $35.1 \pm 0.3$ \\
     & 200M & \textbf{0.652} & \textbf{0.728} & $\mathbf{26.8 \pm 0.3}$ & 0.600 & \textbf{0.774} & $\mathbf{35.1 \pm 0.3}$ \\
    \midrule
    \multirow{7}{*}{Velocityformer broken-$E(3)$} & 0.05M & 0.655 & 0.714 & $24.9 \pm 0.4$ & 0.620 & 0.750 & $31.3 \pm 0.4$ \\
     & 0.2M & \textbf{0.627} & 0.741 & $29.9 \pm 0.4$ & 0.602 & 0.764 & $33.9 \pm 0.4$ \\
     & 0.8M & 0.628 & 0.741 & $30.2 \pm 0.4$ & 0.600 & 0.765 & $34.6 \pm 0.4$ \\
     & 3M & 0.635 & 0.737 & $29.1 \pm 0.4$ & 0.602 & 0.764 & $33.9 \pm 0.4$ \\
     & 12M & 0.635 & 0.736 & $29.4 \pm 0.4$ & 0.602 & 0.763 & $34.3 \pm 0.4$ \\
     & 48M & 0.632 & 0.740 & $30.5 \pm 0.4$ & 0.598 & 0.767 & $35.4 \pm 0.4$ \\
     & 200M & 0.630 & \textbf{0.750} & $\mathbf{30.7 \pm 0.4}$ & \textbf{0.596} & \textbf{0.779} & $\mathbf{35.9 \pm 0.4}$ \\
    \bottomrule
  \end{tabular}
\end{table}

\section{Statistical significance and uncertainties}
\label{app:errors}
The effect of varying the pseudo-random number generator seed was tested on the low-data, 4 simulation boxes setup, for model sizes ranging from 0.8 to 3 million parameters, for 3 distinct random seeds. 
The largest variance among the 3 seeds across model sizes was used as a conservative estimate of the per-model type random seed variance. 
This procedure was chosen as limitations in available compute meant that running the training on multiple seeds for all experimental setups was infeasible.

Variance inherent to the test set was estimated by computing the reported metrics on 5 test sets of 10 simulation boxes each, and computing the standard error on the mean. 
The final uncertainty was calculated by adding the seed variance and test set variance in quadrature. 
The absolute and relative uncertainties for the 4 simulation boxes, 0.8M model category are listed in Table~\ref{tab:errors}. 
Except for $\Delta r$ the relative uncertainties are $\lesssim1\%$. 
The $\Delta r$ uncertainties are somewhat larger due to $\Delta r$ being a ratio of two noisy quantities. 

\begin{table}[ht]
  \centering
  \caption{%
    Estimates of the absolute and relative uncertainties (standard deviations) of the evaluation metrics.%
  }
  \label{tab:errors}
  \small
  \begin{tabular}{l@{\hspace{12pt}}rrr@{\hspace{12pt}}rrr}
    \toprule
    & \multicolumn{3}{c}{Absolute} & \multicolumn{3}{c}{Relative [\%]} \\
    \cmidrule(lr){2-4} \cmidrule(lr){5-7}
    Model & $\sigma_l$ & $\sigma_r$ & $\sigma_{\Delta r}$ & $\sigma_l/l$  & $\sigma_r/r$  & $\sigma_{\Delta r}/\Delta r$  \\
    \midrule
    Transformer & 0.0059 & 0.0074 & 1.29 & 0.88 & 1.07 & 6.07 \\
    GNN & 0.0018 & 0.0010 & 0.19 & 0.27 & 0.14 & 0.76 \\
    U-Net & 0.0011 & 0.0014 & 0.22 & 0.15 & 0.22 & 1.34 \\
    Velocityformer $L_\mathrm{max}{=}0$ & 0.0025 & 0.0026 & 0.45 & 0.34 & 0.41 & 3.92 \\
    Velocityformer $E(3)$ & 0.0014 & 0.0015 & 0.26 & 0.20 & 0.22 & 1.45 \\
    Velocityformer broken-$E(3)$ & 0.0019 & 0.0023 & 0.42 & 0.30 & 0.31 & 1.38 \\
    \bottomrule
  \end{tabular}
\end{table}

\section{Training details and hyperparameters}
\label{app:training}

All models were trained with constant learning rate, the \texttt{AdamW} \citep{Loshchilov2019-adamw} optimiser, and $0.01$ weight decay. 
Removing weight decay, adding learning rate warm-up, and using a cosine learning rate schedule were tested but found to have only negligible effects on our experiments, particularly for \Velformer. 

For the four and 38 simulation boxes training regimes early stopping was employed. 
For the 3800 simulation boxes case training was done for a single epoch. 
The constant learning rate makes the loss comparable across different stages of the single-epoch training, facilitating arguing about scaling behaviour.

Model-specific training details and hyperparameters are described in the following subsections.

The overall compute required for the experiments presented here was about 2000 GPU hours, split between NVIDIA A100 and RTX 4090 GPUs. 

\subsection{\Velformer}
\label{app:velocityformer}
\Velformer largely follows the setup of \citep{Liao2024-equiformerv2}, with modifications to support the broken $E(3)$ symmetry of the galaxy survey data. 
The hyperparameters for the broken-$E(3)$ \Velformer are listed in Table~\ref{tab:velformer-hyperparameters} and use the Equiformer~V2 defaults otherwise. 
For multi-epoch training on four and 38 simulation boxes, we apply a dropout rate of 0.1 and stochastic depth of 0.05.

The embedding of the LOS direction is done by applying the radial embedding to the LOS coordinate, followed by a linear layer that maps to \texttt{sphere\_channels} channels.

We found \Velformer to be remarkably robust to changes in the learning rate, tolerating learning rates up to 0.1 without destabilising training.

\begin{table}
  \centering
  \caption{Hyperparameters of the broken-$E(3)$ \Velformer.}
  \label{tab:velformer-hyperparameters}
  \begin{tabular}{lrrrrrrr}
    \toprule
    & 0.05M & 0.2M & 0.8M & 3M & 12M & 48M & 200M \\
    \midrule
    Parameters & 61.8K & 228K & 790K & 2.94M & 11.5M & 44.5M & 203M \\
    \texttt{batch\_size} & 256 & 128 & 64 & 128 & 64 & 24 & 10 \\
    \texttt{learning\_rate} & $3{\cdot}10^{-3}$ & $3{\cdot}10^{-3}$ & $3{\cdot}10^{-3}$ & $3{\cdot}10^{-3}$ & $10^{-3}$ & $10^{-3}$ & $3{\cdot}10^{-4}$ \\
    \texttt{num\_layers} & 3 & 4 & 4 & 4 & 4 & 6 & 7 \\
    \texttt{lmax\_list} & 1 & 2 & 2 & 4 & 4 & 6 & 6 \\
    \texttt{mmax\_list} & 1 & 2 & 2 & 4 & 4 & 4 & 4 \\
    \texttt{sphere\_channels} & 16 & 16 & 32 & 32 & 64 & 64 & 128 \\
    \texttt{attn\_hidden\_channels} & 8 & 16 & 32 & 32 & 64 & 64 & 128 \\
    \texttt{num\_heads} & 4 & 4 & 4 & 6 & 6 & 8 & 8 \\
    \texttt{attn\_alpha\_channels} & 8 & 16 & 32 & 32 & 64 & 64 & 128 \\
    \texttt{attn\_value\_channels} & 8 & 8 & 16 & 16 & 32 & 32 & 64 \\
    \texttt{ffn\_hidden\_channels} & 8 & 16 & 32 & 32 & 64 & 64 & 128 \\
    \texttt{edge\_channels} & 8 & 16 & 32 & 32 & 64 & 64 & 128 \\
    \texttt{num\_distance\_basis} & 512 & 512 & 512 & 512 & 512 & 512 & 512 \\
    \bottomrule
  \end{tabular}
\end{table}

\subsection{Transformer baseline}
\label{app:tranformer}
The Transformer baseline uses the Transformer implementation from \citet{Brehmer2023-geometric-algebra-transformer}, adapted to accept velocity inputs and to predict velocities. 
The hyperparameters are listed in Table~\ref{tab:transformer-hyperparameters}. 
By default, the model can attend to all galaxies in the input, but it can also work in graph mode by using the graph adjacency matrix as an attention mask. 

\begin{table}
  \centering
  \caption{Hyperparameters of the Transformer baseline.}
  \label{tab:transformer-hyperparameters}
  \begin{tabular}{lrrrrrrr}
    \toprule
    & 0.05M & 0.2M & 0.8M & 3M & 12M & 48M & 200M \\
    \midrule
    Parameters & 50.8K & 200K & 793K & 2.96M & 11.8M & 47.2M & 189M \\
    \texttt{batch\_size} & 2048 & 2048 & 2048 & 1024 & 512 & 256 & 448 \\
    \texttt{learning\_rate} & $3{\cdot}10^{-4}$ & $3{\cdot}10^{-4}$ & $3{\cdot}10^{-4}$ & $3{\cdot}10^{-4}$ & $3{\cdot}10^{-4}$ & $10^{-4}$ & $10^{-4}$ \\
    \texttt{hidden\_channels} & 32 & 64 & 128 & 192 & 384 & 768 & 1536 \\
    \texttt{num\_blocks} & 6 & 6 & 6 & 10 & 10 & 10 & 10 \\
    \texttt{num\_heads} & 8 & 8 & 8 & 8 & 8 & 8 & 8 \\
    \bottomrule
  \end{tabular}
\end{table}

\subsection{GNN baseline}
\label{app:gnn-baseline}
The GNN baseline uses the implementation from \citet{Huang2025-cosmobench}. 
While that implementation already supports the velocity reconstruction task, we adapt it to accept the linear velocity estimate as input feature. 
The hyperparameters are listed in Table~\ref{tab:gnn-hyperparameters}. 

The GNN was found to be unstable at large model sizes and with fully connected graphs. 
We therefore do not report results for the 200M parameter model and note that the fully connected graph variant (see Appendix~\ref{app:varyk}) likely underperforms due to training instability rather than a fundamental limitation of the architecture.

\begin{table}
  \centering
  \caption{Hyperparameters of the GNN baseline.}
  \label{tab:gnn-hyperparameters}
  \begin{tabular}{lrrrrrrr}
    \toprule
    & 0.05M & 0.2M & 0.8M & 3M & 12M & 48M & 200M \\
    \midrule
    Parameters & 64.0K & 189K & 747K & 2.97M & 12.3M & 49.2M & 168M \\
    \texttt{batch\_size} & 2048 & 1024 & 640 & 320 & 96 & 120 & 88 \\
    \texttt{learning\_rate} & $10^{-3}$ & $10^{-3}$ & $10^{-3}$ & $10^{-3}$ & $3{\cdot}10^{-4}$ & $3{\cdot}10^{-4}$ & $10^{-4}$ \\
    \texttt{d\_hidden} & 64 & 96 & 192 & 384 & 640 & 1280 & 2048 \\
    \texttt{message\_passing\_steps} & 3 & 4 & 4 & 4 & 6 & 6 & 8 \\
    \bottomrule
  \end{tabular}
\end{table}

\subsection{U-Net baseline}
\label{app:unet}
Following previous work~\citep[e.g.][]{Ganeshaiah-Veena2023-unet-lpt-vel-recon,Wu2023-unet-vel-recon,Wang2024-ml-vel-recon,Xiao2025-ml-vel-recon}, we include a grid-based U-Net~\citep{Ronneberger2015-unet} as a baseline. 
Results are reported in the main text; this appendix describes the implementation.

The input point cloud, consisting of a varying number of galaxy positions and associated linear velocity estimates, is painted onto a $64^3$ grid using cloud-in-cell (CIC) assignment, producing a four-channel grid (one density channel and three velocity channels). 
The grid resolution is fixed at $64^3$; although the galaxy fields are sparse ($\mathcal{O}(100)$ galaxies per $71.4\,\hMpc$ box), we found that reducing the resolution to $32^3$ degraded performance, while increasing it to $128^3$ did not yield significant improvements while greatly increasing computational cost.

The architecture is a standard 3D U-Net with double convolution blocks using $3{\times}3{\times}3$ kernels, max-pool downsampling, transposed convolution upsampling, and skip connections between the symmetric encoder and decoder. 
Being fully convolutional, it is translationally equivariant but not rotationally equivariant. 
After the U-Net, the predicted velocity field is sampled back to particle positions via CIC interpolation. 
The network includes an additional residual connection at the particle level: the linear velocity estimates are added directly to the interpolated output, bypassing the grid representation entirely. 
This differs from the internal U-Net skip connections, which operate on grid features, and outperformed using the velocity estimates only at the input.
We train four model variants with hyperparameters listed in Table~\ref{tab:unet-hyperparameters}.

\begin{table}
  \centering
  \caption{Hyperparameters of the U-Net baseline.}
  \label{tab:unet-hyperparameters}
  \begin{tabular}{lrrrr}
    \toprule
    & 0.8M & 3M & 12M & 48M \\
    \midrule
    Parameters & 790K & 3.15M & 12.6M & 50.4M \\
    \texttt{batch\_size} & 64 & 32 & 8 & 12 \\
    \texttt{learning\_rate} & $3{\cdot}10^{-4}$ & $3{\cdot}10^{-4}$ & $3{\cdot}10^{-4}$ & $3{\cdot}10^{-4}$ \\
    \texttt{grid\_size} & $64^3$ & $64^3$ & $64^3$ & $64^3$ \\
    \texttt{encoder\_features} & [12,24,48,96] & [24,48,96,192] & [48,96,192,384] & [96,192,384,768] \\
    \bottomrule
  \end{tabular}
\end{table}

\section{Data details}
\label{app:ablations}

\subsection{Graph construction}
\label{app:varyk}
As the graph structure is an arbitrary choice informed by implementation limitations rather than physical principles, we test the effect of varying the number of nearest neighbours $k$ used to construct the graph. 
The test was performed for the mid-data, 38 simulation boxes setup, for the 0.8M parameter variants of the Transformer, GNN, and the broken-$E(3)$ \Velformer. 
To limit the connectivity for the Transformer we use the graph adjacency matrix as an attention mask. 

The MSE $l$, correlation coefficient $r$, and improvement $\Delta r$ for different connectivities are listed in Table~\ref{tab:vary_k}. 
Increasing the connectivity generally improves performance, with the fully connected graph performing best for the Transformer and \Velformer. 
We hypothesise that the same would be true for the GNN if it were not for unstable training dynamics. 
This increased performance comes at prohibitive computational cost however, especially for \Velformer, such that we do not further explore fully connected graphs in this work. 

\begin{figure}
  \centering
  \includegraphics[width=\textwidth]{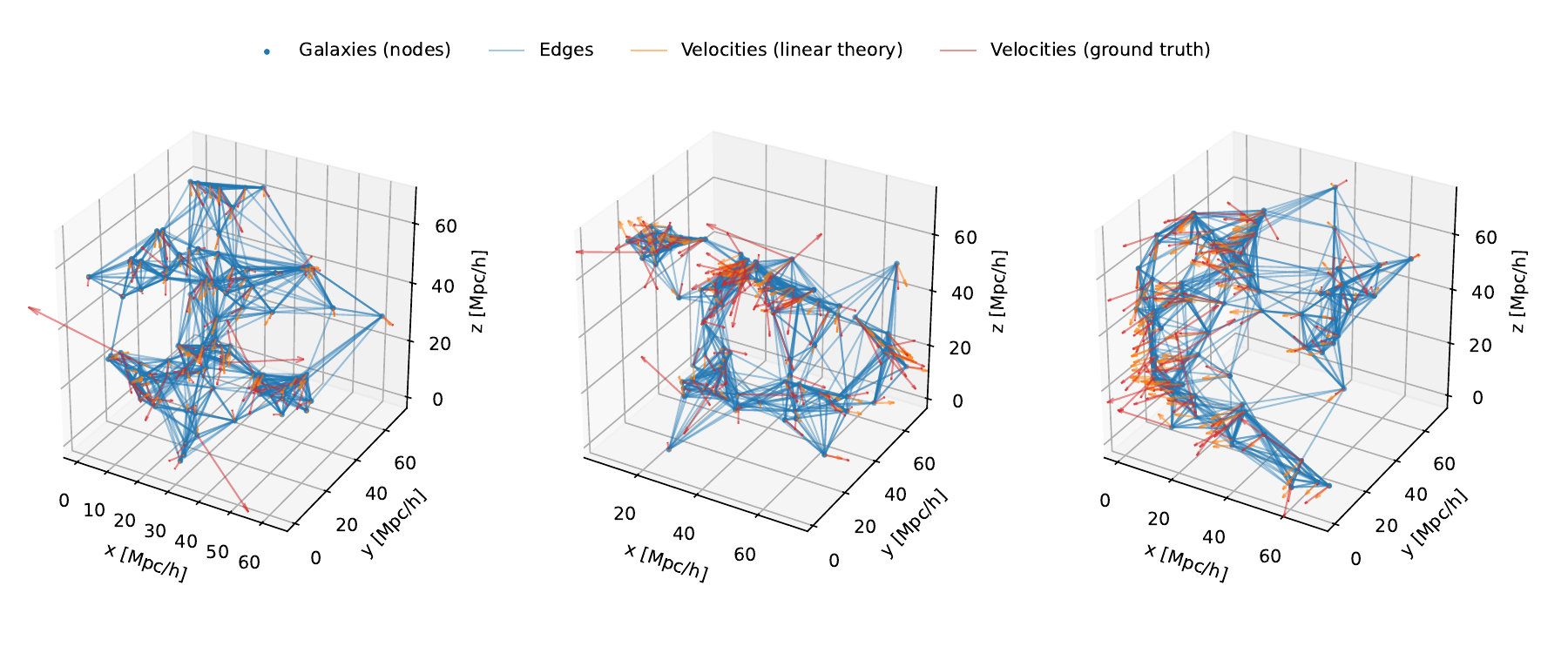}
  \caption{%
    Sample galaxy point clouds from the validation set with the imposed graph structure, input linear theory velocities, and true velocities.%
  }
  \label{fig:sample_graphs}
\end{figure}

Three sample galaxy point clouds with the imposed graph structures are visualised in Fig.~\ref{fig:sample_graphs}.

\begin{table}[ht]
  \centering
  \caption{%
    Velocity MSE $l$, correlation coefficient $r$, and improvement $\Delta r$ for
    different graph connectivity (\knn vs.\ fully connected)
    at fixed model size.
    The best value per metric is boldfaced within each model group.%
  }
  \label{tab:vary_k}
  \small
  \begin{tabular}{ll@{\hspace{10pt}}rrr}
    \toprule
    Model & Connectivity & $\downarrow l$ & $\uparrow r$ & \multicolumn{1}{l}{\hspace{6pt}$\uparrow \Delta r$ [\%]} \\
    \midrule
    \multirow{3}{*}{Transformer} & $k{=}10$ & 0.670 & 0.707 & 24.3 \\
     & $k{=}20$ & 0.657 & 0.718 & 26.2 \\
     & Fully connected & \textbf{0.629} & \textbf{0.740} & \textbf{29.1} \\
    \midrule
    \multirow{3}{*}{GNN} & $k{=}10$ & 0.646 & 0.733 & 28.2 \\
     & $k{=}20$ & \textbf{0.641} & \textbf{0.739} & \textbf{29.6} \\
     & Fully connected & 0.656 & 0.734 & 29.5 \\
    \midrule
    \multirow{3}{*}{Velocityformer broken-$E(3)$} & $k{=}10$ & 0.602 & 0.763 & 34.2 \\
     & $k{=}20$ & 0.594 & 0.770 & 35.5 \\
     & Fully connected & \textbf{0.590} & \textbf{0.774} & \textbf{37.0} \\
    \bottomrule
  \end{tabular}
\end{table}

\subsection{Varying $\nsplit$}
\label{app:varynsplit}
The factorisation of the velocity reconstruction task into large and small scales assumes that the linear velocity estimate captures the velocity field on scales larger than the sub-box size, while the galaxy positions within the sub-box are sufficient to reconstruct the velocity field on smaller scales.
Choosing the sub-boxes to be too large defeats the efficiency gains of the factorisation, while choosing them to be too small may lead to insufficient information within the sub-box to reconstruct the small-scale velocity field.

We test the effect of varying the sub-box partition size $\nsplit$ for the broken-$E(3)$ \Velformer at 0.8M parameters, for both the 4 and 38 simulation boxes setups.
The results are listed in Table~\ref{tab:vary_nsplit}. 
The performance is relatively stable across the range of $\nsplit$ tested, with a slight preference for larger sub-boxes. 
We chose $\nsplit{=}14$ for the main results as it is the smallest sub-box size that does not show a clear performance degradation, thus maximising the number of training samples we can construct from one simulation box.

\begin{table}[ht]
  \centering
  \caption{%
    Effect of sub-box partition size ($n_\mathrm{split}$) on velocity MSE $l$, correlation coefficient $r$, and improvement $\Delta r$.
    Results are for \Velformer broken-$E(3)$ at $0.8\,\mathrm{M}$ parameters.%
  }
  \label{tab:vary_nsplit}
  \small
  \begin{tabular}{r@{\hspace{10pt}}rrr@{\hspace{10pt}}rrr}
    \toprule
    $n_\mathrm{split}$ & \multicolumn{3}{c}{Low-data} & \multicolumn{3}{c}{Mid-data} \\
    \cmidrule(lr){2-4} \cmidrule(lr){5-7}
    & $\downarrow l$ & $\uparrow r$ & \multicolumn{1}{l}{\hspace{6pt}$\uparrow \Delta r$ [\%]} & $\downarrow l$ & $\uparrow r$ & \multicolumn{1}{l}{\hspace{6pt}$\uparrow \Delta r$ [\%]} \\
    \midrule
    8 & 0.653 & 0.716 & 25.2 & 0.599 & 0.767 & 34.1 \\
    10 & 0.638 & 0.731 & 28.0 & 0.599 & 0.767 & 34.2 \\
    12 & 0.630 & 0.739 & 29.7 & 0.599 & 0.767 & 34.6 \\
    14 & 0.631 & 0.738 & 29.7 & 0.603 & 0.763 & 34.2 \\
    16 & 0.631 & 0.737 & 30.0 & 0.606 & 0.759 & 34.0 \\
    \bottomrule
  \end{tabular}
\end{table}

\subsection{Zero-shot data}
\label{app:zero-shot-setup}
Here we briefly describe the datasets used for testing the generalisation performance. 
Key properties of the galaxy samples, the galaxy number density and linear galaxy bias, are listed in Table~\ref{tab:sample_properties}.

\begin{table}[ht]
  \centering
  \caption{%
    Number density and linear galaxy bias for the different simulation sets considered for zero-shot generalisation.%
  }
  \label{tab:sample_properties}
  \begin{tabular}{lrrrr}
    \toprule
    & Fiducial & A & B & \flamingo \\
    \midrule
    Number density [$10^4\ihMpcV$] & 3.55 & 3.59 & 3.07 & 3.5 \\
    Linear bias & 2.08 & 2.06 & 1.92 & 2.00 \\
    \bottomrule
  \end{tabular}
\end{table}

\paragraph{Cosmology}
We select two cosmologies ($i=1789$, `A' and $i=816$, `B') from the \Quijote\texttt{nwLH} latin hypercube set. 
The parameters are listed in Table~\ref{tab:quijote-cosmologies}. 
The shift of the two cosmologies with respect to the fiducial set that was used for training is roughly consistent with the uncertainty on cosmology from DESI \citep{DESI2025-dr1-full-shape,DESI2025-dr2-bao-cosmology}. 
\Velformer can thus generalise across our uncertainty on the true cosmological parameters of the Universe. 

\begin{table}[ht]
  \centering
  \caption{%
    Cosmological parameters for the \Quijote simulation sets.%
  }
  \label{tab:quijote-cosmologies}
  \begin{tabular}{lrrrrrrr}
    \toprule
     & $\Omega_\mathrm{m}$ & $\Omega_\mathrm{b}$ & $h$ & $n_\mathrm{s}$ & $\sigma_8$ & $m_\nu$ [eV] & $w$ \\
    \midrule
    Fiducial & 0.3175 & 0.049 & 0.6711 & 0.9624 & 0.834 & 0.0 & -1.0 \\
    A & 0.3283 & 0.03661 & 0.5989 & 0.9681 & 0.8465 & 0.0686575 & -1.01515 \\
    B & 0.2893 & 0.05521 & 0.5697 & 0.9105 & 0.8617 & 0.0607375 & -1.02625 \\
    \flamingo & 0.316 & 0.0494 & 0.673 & 0.966 & 0.812 & 0.06 & -1.0 \\
    \bottomrule
  \end{tabular}
\end{table}

\paragraph{\flamingo galaxy catalogue}
The \flamingo sample is built from the \texttt{L1\_m9} simulation, \texttt{Planck} variant, and \texttt{exclusive\_sphere\_300kpc} halo catalogue for the $z{=}0.5$ snapshot. 
For details on these data product, see \citet{Helly2026-flamingo-data-release} and \url{https://flamingo.strw.leidenuniv.nl/}.
We perform a simple cut on stellar mass to reduce the number density to $3.5\times 10^{-4}\ihMpcV$. 
Positions and velocities are those of the stellar centre of mass.



\end{document}